\documentclass[RNAAS,twocolumn]{aastex62}
\usepackage{graphicx,url,amssymb,amsmath,color,units,wasysym,epsfig,epstopdf,enumerate,tabularx,subfigure,hyperref}
 
\usepackage{newtxtext,newtxmath}
\usepackage[T1]{fontenc}
\usepackage{ae,aecompl}

\usepackage[normalem]{ulem}
\usepackage[bottom]{footmisc}

\renewcommand{\sc}[1]{\textsc{#1}}

\newcommand{\new}[1]{{#1}}

\colorlet{darkorange}{black!30!orange!70!}
\newcommand{\postpnp}[1]{{#1}}

\newcommand{\SPA}{School of Physics and Astronomy, Monash University, Clayton VIC 3800, Australia}
\newcommand{\OzGravMonash}{OzGrav: The ARC Centre of Excellence for Gravitational Wave Discovery, Clayton VIC 3800, Australia}

\begin{document}

\title{Signs of eccentricity in two gravitational-wave signals may indicate a sub-population of dynamically assembled binary black holes}

\author{Isobel Romero-Shaw}
    \email{isobel.romero-shaw@monash.edu}
\affiliation{\SPA}
\affiliation{\OzGravMonash}

\author{Paul D. Lasky}
\affiliation{\SPA}
\affiliation{\OzGravMonash}

\author{Eric Thrane}
\affiliation{\SPA}
\affiliation{\OzGravMonash}

\begin{abstract}
The orbital eccentricity of a merging binary black hole leaves an imprint on the associated gravitational-wave signal that can reveal whether the binary formed in isolation or in a dynamical environment, such as the core of a dense star cluster.
We present measurements of the eccentricity of 26 binary black hole mergers in the second LIGO--Virgo gravitational-wave transient catalog, updating the total number of binary black holes analysed for orbital eccentricity to 36.
Using the \texttt{SEOBNRE} waveform, we find the data for GW190620A is poorly explained by the zero-eccentricity hypothesis (frequentist $p$-value $\lesssim 0.1\%$).
\new{Using a log-uniform prior on eccentricity, the eccentricity at $\unit[10]{Hz}$ for GW190620A is constrained to $e_{10}\geq0.05$ ($0.1$) at $74\%$ ($65\%$) credibility.
With this log-uniform prior, we obtain a $90\%$ credible lower eccentricity limit of $0.001$, while assuming a uniform prior leads the data to prefer $e_{10} \geq 0.11$ at $90\%$ credibility.}
This is the second measurement of a binary black hole system with statistical support for non-zero eccentricity; the intermediate-mass black hole merger GW190521 was the first. 
Interpretation of these two events is currently complicated by waveform systematics; we are unable to simultaneously model the effects of relativistic precession and eccentricity.
However, if these two events are, in fact, eccentric mergers, then there are potentially many more dynamically assembled mergers in the LIGO--Virgo catalog without measurable eccentricity; $\gtrsim 27\%$ of the observed LIGO--Virgo binaries may have been assembled dynamically in dense stellar environments ($95\%$ credibility).
\end{abstract}

\section{Introduction}
\label{sec:intro}

The second gravitational-wave transient catalog~\citep[GWTC-2;][]{GWTC-2} of the LIGO--Virgo collaboration~\citep{ALIGO,AdvancedVirgo} confirmed the detection of 36 new binary black hole (BBH) mergers. 
Combined with the mergers presented in the first catalog \citep[GWTC-1;][]{GWTC-1}, there are now 46 confirmed BBH merger detections.\footnote{The exact number of ``confirmed'' mergers depends on the choice of detection threshold. Using a stricter threshold \cite{GWTC-2_RnP} counts 44 confirmed BBH mergers.} 
This abundance of events poses an intriguing question in gravitational-wave astronomy: how did these merging binaries form? 

There are two primary channels that can produce binary compact object mergers that can merge in a Hubble time: isolated evolution and dynamical formation. 
An isolated binary contains two stars that are born together and evolve together, undergoing some mechanism that allows the two components to become close to merge within the age of the Universe, without merging before they become compact objects.
A variety of mechanisms have been proposed, including common envelope evolution \citep[e.g.,][]{Livio88, Bethe98, Ivanova13, Kruckow16}, chemically homogeneous evolution \citep[e.g.,][]{deMink10, deMink16, Marchant16}, stable mass accretion onto a black hole from its stellar companion \citep{Heuvel17,Neijssel19,Bavera21}, or ambient gas-driven fallback \citep[e.g.,][]{Tagawa18}.
In contrast, a dynamically {assembled} binary does not become bound until the two components have already evolved into compact objects. 
This can occur in places like globular~\citep[e.g.,][]{Rodriguez16,Samsing17,Hong18} and nuclear~\citep{Grishin:2018:triples,Fragione:2019:BH-NS-AGN,Hoang2018,Fragione2020:nuclear} star clusters. 
In these dense environments, mass segregation leads to a dark compact object core, where objects can undergo many frequent gravitational interactions~\citep[e.g.,][]{Wen02, Antonini15, Morscher15, Rodriguez18a, Rodriguez18b, Samsing18, DRAGON}.
Subsequently, black holes can form binaries that are hardened through interactions with other compact objects, eventually merging.

There are three \postpnp{intrinsic properties of a binary that can distinguish its formation channel}: its component masses, component spins, and orbital eccentricity.
Multiple studies have shown that these properties can be used to identify the formation channel of a single binary and to constrain the relative fraction of mergers contributed by that channel to the overall merger rate~\citep[e.g.,][]{Vitale15, Farr17, Zevin:2020:channels, Zevin:2021:seleccentricity,TalbotThrane17}.
\postpnp{The formation channels of populations of mergers can also be distinguished using the redshift evolution of the merger rate~\citep[e.g.,][]{RodriguezLoeb2018,Ng:2021:redshift}; however, it will take upwards of $\sim100$ detections for this to become possible~\citep{Fishbach:2018:mergerrate}.}

Identifying mergers with component masses between $\sim60$--$130$~M$_{\odot}$ may indicate the presence of hierarchical mergers (from repeated dynamical mergers)~\citep[e.g.,][]{Fishbach17,Kimball:2020:hierarchical,Kimball2021}.
As pair-instability and pulsational pair-instability supernovae enforce an upper limit on the mass of a black hole that can form through stellar collapse~\citep{HegerWoosley02, Ozel10, Belczynski16, Marchant16, Fishbach17, Woosley17, TalbotThrane18}, there is thought to be a dearth of black holes in this range, although these boundaries are sensitive to assumptions about the underlying physics~\citep[see, e.g.,][and references within]{Sakstein2020, Belczynski2020,Farmer2019,Farmer2020}.
In dynamical environments, on the other hand, merger remnants may go on to merge again if their formation kick does not eject them from the cluster, leading to black holes within this mass gap~\citep{Gerosa17, Rodriguez19, Bouffanais19, fragione20, Samsing:2020:massgap, Kremer:2020:massgap, Kimball:2020:hierarchical}.
The intermediate-mass black hole binary GW190521 \citep{GW190521-detection} has been interpreted as such a hierarchical merger~\citep[e.g.,][]{Kimball2021, Fragione2020:GW190521, Anagnostou:2020:GW190521}.
For an alternative interpretation, see \cite{Nitz2020,Olsen2020}, which argue that GW190521 may be an intermediate-mass ratio inspiral with $q\equiv m_2/m_1 \approx 0.09$.
In this work we assume the currently conventional interpretation, that $q\approx0.8$.

Observing a population of BBH events in which some fraction of binaries have black-hole spins anti-aligned with the orbital angular momentum would also hint that dynamical formation is at play~\citep{ Stevenson17dlk, TalbotThrane17, GWTC-2_RnP}.
Binary stars evolving together in the field tidally interact, leading them to have preferentially aligned spins~\citep[e.g.,][]{Gerosa:2018wbw,Kalogera2000, Campanelli06}. 
While the supernovae of one object can lead to a slight change in the spin orientation of the other, this change is believed to be minor~\citep[see, e.g.,][and references therein]{OShaughnessy17, Gerosa:2018wbw}.
In contrast, objects that become bound during a gravitational interaction in the core of a dense star cluster may have any spin orientation relative to each other, and so we expect a population of binaries formed in clusters to have an isotropic spin distribution~\citep{Rodriguez16}.

The LIGO--Virgo analysis of GWTC-2 found evidence for anti-aligned spin in the detected BBH population, and inferred from this that $\approx25-93\%$ of the observed BBH had formed dynamically, at $90\%$ credibility~\citep{GWTC-2_RnP}.
However, \citet{Roulet2021} dispute this, finding that the signature from \cite{GWTC-2_RnP} is a model-dependent artefact.
In either case, the dynamical formation scenario is unlikely to produce the entirety of mergers observed by LIGO and Virgo.
The presence of $\approx 10$ BBH signals with black-hole spins preferentially aligned with the orbital angular momentum suggests $\gtrsim23\%$ of BBH events are associated with field mergers.

The third \postpnp{intrinsic} property of a binary that can act as a signature of dynamical formation is its orbital eccentricity close to merger.
Gravitational-wave emission efficiently circularises binaries on a shorter timescale than they tighten~\citep{Peters64,Hinder07}.
We thus expect negligible eccentricity in the orbits of field binaries at detection---excepting field triples, a topic we return to below.
In a dynamical environments such as dense star clusters, however, binaries can be driven to merge rapidly.
They do not always have time to radiate away their eccentricity before they merge, and so they may retain detectable eccentricity when {their gravitational radiation enters} the LIGO--Virgo band at {gravitational-wave} frequencies $\gtrsim\unit[10]{Hz}$~\citep{Morscher15, Samsing17, Rodriguez18a, Rodriguez18b, Gondan18, Zevin18}.\footnote{{Throughout, we use the word “frequency” to refer to gravitational-wave frequency as opposed to orbital frequency.}}

Dense star clusters are arguably the most well-studied dynamical formation environment~\citep[see, e.g.,][]{Sigurdsson93, PortegiesZwart99, OLeary05, Samsing13, Morscher15, Gondan17, Samsing17, Rodriguez18b, Randall17, Randall18, SamsingDOrazio18, Samsing18, Rodriguez18a, Fragione18, Fragione19b, Bouffanais19}.
Simulations of compact binary formation in such environments lead us to expect that $\sim5\%$ ($\sim7\%$) of their binary black holes retain eccentricities $e_{10} \geq 0.1 (0.05)$ when their gravitational radiation frequency reaches \unit[10]{Hz};~see~\citet{Samsing13, Samsing17, Rodriguez18a, Rodriguez18b, Samsing18, Kyle_data_paper, Zevin:2021:seleccentricity} and references within.
Observing eccentricity in the gravitational waveform of a BBH coalescence therefore indicates that the system was formed dynamically.
Young open clusters have also been proposed as a competitive channel~\citep[e.g.,][]{Giacomo2021}, which may produce mergers with traits associated with either dynamical or isolated formation.

Further alternatives to dynamical formation in dense star clusters include dynamical formation in active galactic nuclei discs~\citep{Yang2019, McKernan2020, Grobner20, Li2021}, which may be efficient factories for eccentric binary black holes~\citep{Samsing:2020:AGN, Tagawa:2021:AGN}.
However, the distribution of mass, spin, and eccentricity for binary black holes in active galactic nuclei discs are comparatively poorly understood owing to the complicated environment.

{Additional classes of formation mechanism include} field triples~\citep{Silsbee16, Antonini17, Fishbach17a, Rodriguez18jqu, triplespin, Liu19} and quadruples~\citep{quadruples, quadruples2}, which can cause the spins and eccentricities of isolated binary mergers to somewhat resemble those of dynamical mergers.
Field triple mergers can have high eccentricities, as the third component can drive up the eccentricity of the inner binary in a process known as Kozai-Lidov resonance~\citep{Kozai62, Lidov62}.
The rate of mergers driven by Kozai-Lidov resonance in the field is thought to be low, unless black hole natal kicks are small and the formation metallicities of the systems are low~\citep{Silsbee16, Antonini17, Rodriguez18jqu, triplespin, Liu19}.

{It has also been suggested that the observed population of mergers may contain primordial black holes, which can have lower and/or higher masses than those formed through stellar collapse~\citep[e.g.,][]{Bird:2016:PBH, Sasaki:2016:PBH,AliHamoud:2017:PBH,Franciolini:2021:PBH,DeLuca:2021:PBH,Chen:2021:PBH}. 
However, there is at present no evidence for the existence of primordial black holes~\citep{Carr:2020:PBH}, and if they do exist, it is not clear that they form merging binaries; \cite[see, e.g.,][]{Korol}}.

In \cite{Romero-Shaw:2019itr}, we presented measurements of orbital eccentricity for BBH events in GWTC-1, constraining the eccentricity of these ten mergers to less than $0.1$ at $10$~Hz.
\postpnp{This result was in agreement with that of ~\citet{EccentricCWB19}, which found no eccentric signals within the data from LIGO and Virgo's first and second observing runs.}
In \citet{Romero-Shaw:2020:GW190521}, we presented tentative evidence that the highest-mass binary so far detected in gravitational waves, GW190521A~\citep{GW190521-detection, GW190521-implications}\footnote{In this paper, we use the short event name, \postpnp{appending an A (B) if the event is the first (second) on that date}.
}, had non-zero eccentricity, although the purported signal could also be the result of general relativistic precession induced by black-hole spin.  
This conclusion was supported by \citet{Gayathri2020}.
GW190521A may therefore be the first observation of an eccentric binary in the population of LIGO-Virgo detected events.

In this work, we present measurements of eccentricity for $36$ of the $46$ BBHs in GWTC-2.\footnote{{The events in the GWTC-2 catalogue were detected using quasi-circular waveform templates.
Some events were also detected with ``burst'' pipelines using excess power techniques \citep[e.g.,][]{Cornish:2014kda,Coughlin:2015:detect,Drago:2020:CWB}.
As eccentricity grows, signals increasingly deviate from quasi-circular signal templates, so can appear with low significance in circular searches~\citep[e.g.,][]{Brown09}.
Unmodeled analyses can be particularly powerful in this case~\citep[e.g.,][]{Dalya:2021:BW_reconstructed_signals}, and eccentric signals may be recovered with a higher signal-to-noise ratio in a burst search than a circular search.
All GWTC-2 candidates were detected with at least one circular search pipeline.
 }}
We highlight GW190620A, an event for which the $e_{10} \geq 0.1$ hypothesis is preferred to the $e_{10}<0.1$ case by a Bayes factor of $\mathcal{B} = 18.6$.
We detail our analysis method in Section \ref{sec:methods}, where we provide updates to the analysis methods used in \citet{Romero-Shaw:2019itr, Romero-Shaw:2020:GW190425,  Romero-Shaw:2020:GW190521}.
Our results are presented in Section \ref{sec:results}, where we investigate events that have significant posterior support for $e_{10} \geq 0.05$.
We discuss the broader astrophysical interpretation of our results in Section \ref{sec:discussion}.

\section{Method}\label{sec:methods}

\begin{figure*}
    \centering
    \includegraphics[width=0.9\textwidth]{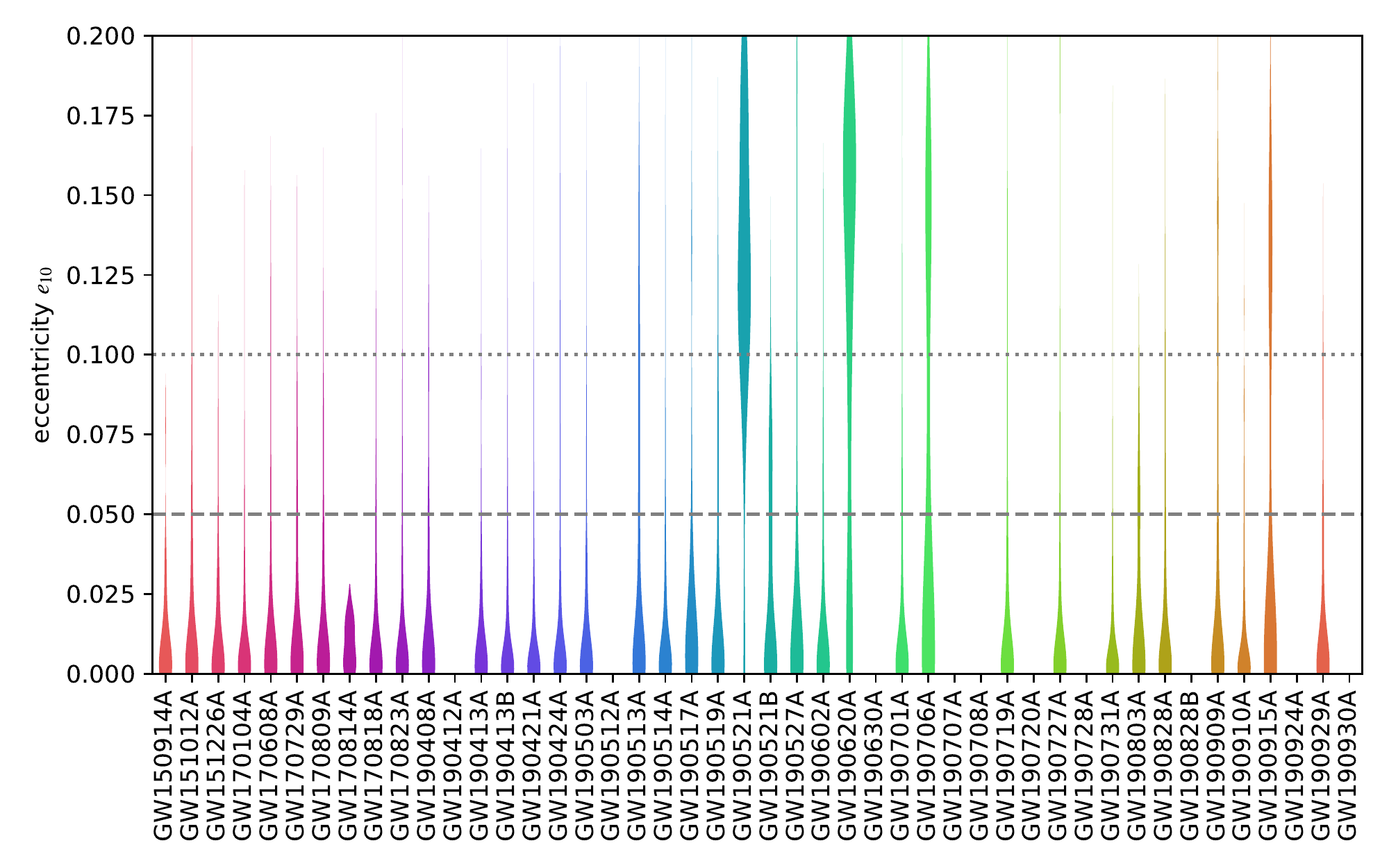}
    \caption{Marginal posterior distributions on eccentricity at $\unit[10]{Hz}$, $e_{10}$, for 36 binary black hole merger events in GWTC-2 \citep{GWTC-2}. 
    We assume a log-uniform eccentricity prior, which is suitable when we do not know the order of magnitude for $e_{10}$.
    The ten low-mass events that require further analysis due to under-sampling are left blank. 
    For each event, the width of the violin at each value of eccentricity is proportional to the posterior distribution at that value. 
    Eccentricity posteriors for events in GWTC-1 and for GW190521A were originally presented in \citet{Romero-Shaw:2019itr} and \citet{Romero-Shaw:2020:GW190521}, respectively.
    {These previously-reported results have here been reweighted from their original prior on eccentricity, which was log-uniform between $10^{-6}$ and $0.2$, to the prior used for analysing the new GWTC-2 events, which is log-uniform between $10^{-4}$ and $0.2$}.}
    \label{fig:eccentricityviolin}
\end{figure*}

We use the likelihood reweighting (importance sampling) method described in \citet{Romero-Shaw:2019itr}, inspired by the importance sampling method used in \citet{Payne2019}, to efficiently estimate sets of posterior distributions for eccentricity.
This method has been tested using injection studies~\citep{Romero-Shaw:2019itr, Romero-Shaw:2020:GW190521}
to correctly recover the injected eccentricity of injected aligned-spin signals.
We obtain initial samples using a quasi-circular waveform model \texttt{IMRPhenomD}~\citep{Khan15} for our proposal likelihood.
These samples are then reweighted using eccentric waveform model \texttt{SEOBNRE}~\citep{SEOBNRE,validationSEOBNRE} to obtain samples from our target distribution. 
We perform Bayesian inference using \texttt{bilby} and the \texttt{bilby\_pipe} pipeline~\citep{bilby, Romero-Shaw:2020:Bilby}, running five parallel analyses with unique seeds for each event. 
We analyse publicly-available data from GWTC-2~\citep{GWOSC:GWTC-2}, using a combination of the LIGO Livingston, LIGO Hanford and Virgo detectors that is consistent with the LIGO-Virgo analysis for each event.

We use power spectral densities generated using \texttt{BayesWave} \citep{Littenberg:2014oda}.
{We do not factor calibration uncertainty into our analysis; errors on our results caused by neglecting calibration uncertainty are expected to be negligible \citep[e.g.,][]{Payne:2020:calibration,Vitale:2021:calibration}. 
Similarly, we do not marginalise over the uncertainty in the noise power spectral density, but marginalising over this uncertainty is expected to yield modest changes in the posterior widths of $\lesssim 5\%$ \citep{Biscoveanu:2020:PSD}.}

Our sampling and reference frequencies are $4096$~Hz and $10$~Hz, respectively.
We use $20$~Hz as the default minimum frequency of analysis in all detectors for all newly-analysed events, except for GW190727A, which has a minimum frequency of $50$~Hz in the LIGO Livingston detector in the LIGO-Virgo analysis~\citep{GWTC-2}.\footnote{{\texttt{SEOBNRE} is defined such that the minimum frequency requested in the waveform is also the reference frequency for the eccentricity. We therefore generate waveforms from \unit[10]{Hz}, but only use the frequency content from \unit[20]{Hz} and above in our analyses.}}
We use the \texttt{dynesty}~\citep{dynesty} sampler with $1000$ live points, $100$ walks and $10$ auto-correlation times.
To avoid spectral leakage, we soften the abrupt start of the time-domain inspiral using a half-Tukey window that turns on over $\unit[0.5]{s}$.

We use standard priors for extrinsic angle parameters.
We use a prior on luminosity distance $d_{\rm L}$ that is uniform in the source-frame.
Our prior on mass ratio $q$ is uniform between $0.125$ and $1$, where the lower bound is restricted by the choice of waveform approximants.
{The prior on the $\hat z$ component of the black hole spin vectors $\chi^z_i$ is created by combining a uniform prior on the component spin magnitudes, $\chi_i$, with an isotropic prior for the spin orientation.
Each $\chi_i$ is capped at $0.6$, as \texttt{SEOBNRE} cannot tolerate spins of greater magnitude than this.
This creates a prior with limits at $\chi_i = \pm 0.6$ and a peak at $\chi_i = 0$.}
We adopt a uniform prior on chirp mass $\mathcal{M}$.

The reweighting procedure is near-identical to that used in \citet{Romero-Shaw:2020:GW190521}, which built on that described in \citet{Romero-Shaw:2019itr}, except that we increase the lower bound on our prior for $e_{10}$ to $e_{10}=10^{-4}$ since we cannot resolve the eccentricity for signals below this point.
We employ a log-uniform prior for eccentricity, which is suitable given that we are unsure about the order of magnitude for $e_{10}$.
{For completeness, we also provide results obtained under a uniform eccentricity prior over the same range. The probability distributions over eccentricity (obtained by dividing out the log-uniform prior in post-processing) are presented in Figure \ref{fig:all_eccentric_likelihoods} in Appendix~\ref{sec:likelihoodApp}.}

\begin{table*}
\centering
\caption{A summary of the eccentricity signature for the 12 events with the most support for $e_{10} \geq 0.05$. 
The second and third columns provide the \postpnp{percentage of} posterior support for $e_{10}>0.1$ and $e>0.05$ respectively.
\postpnp{These two values are typical used as thresholds for `detectable' binary eccentricity at \unit[10]{Hz} using operational gravitational-wave detectors~\citep[e.g.,][]{Lower18,Samsing17, Samsing:2018:MOCCA1,Rodriguez18a,Rodriguez18b,Zevin18,Zevin:2021:seleccentricity}, although the true threshold for eccentricity sensitivity is unique to each signal.}
The next two columns provide the natural log Bayes factors $\ln \mathcal{B}$ for the hypotheses that $e_{10} \geq 0.1$ ($0.05$) against the hypothesis that $e_{10} < 0.1$ ($0.05$).
The two most compelling candidates for eccentric mergers are highlighted in bold.
These same parameters for other events in GWTC-2 are provided in {Appendix~\ref{sec:quasicircularApp}}.
\label{tab:Bayes_and_percentages}}
\bgroup
\def\arraystretch{1.5}
\begin{tabular}{c|c|c|c|c|c}
Event name & percentage $e_{10} \geq 0.1$ & percentage $e_{10} \geq 0.05$ & $\ln \mathcal{B} (e_{10} \geq 0.1)$  & $\ln \mathcal{B} (e_{10} \geq 0.05)$ & reweighting efficiency (\%) \\
\hline
GW190424A & $8.12$ & $17.09$ & $-0.11$ & $-0.08$ & 85 \\
GW190513A & $13.28$ & $27.33$ & $0.45$ & $0.53$ & 49 \\
\textbf{GW190521A} & $\mathbf{92.25}$ & $\mathbf{93.42}$ & $\mathbf{4.65}$ & $\mathbf{3.90}$ & $\mathbf{2}$ \\
GW190521B & $2.43$ & $23.21$ & $-0.17$ & $0.55$ & 6 \\
GW190527A & $8.64$ & $17.72$ & $-0.07$ & $-0.06$ & 15 \\
\textbf{GW190620A} & $\mathbf{65.72}$ & $\mathbf{74.27}$ & $\mathbf{2.90}$ & $\mathbf{2.48}$ & $\mathbf{10}$ \\
GW190706A & $28.27$ & $38.02$ & $1.36$ & $1.01$ & 42 \\
GW190719A & $9.29$ & $19.01$ & $0.04$ & $0.07$ & 70 \\ 
GW190727A & $8.27$ & $17.07$ & $-0.14$ & $-0.09$ & 87 \\
GW190828A & $7.30$ & $19.37$ & $-0.18$ & $0.10$ & 48 \\
GW190909A & $15.61$ & $25.91$ & $0.60$ & $0.44$ & 86 \\
GW190915A & $21.60$ & $33.35$ & $0.99$ & $0.77$ & 9 \\
\end{tabular}
\egroup
\end{table*}

Like other eccentric waveform models~\citep[e.g.,][]{eccentricFD,TEOBResumS}, \texttt{SEOBNRE} does not include a variable mean anomaly.
The phase modulations caused by a varying mean anomaly cannot be fully accounted for by reference phase- and time-marginalisation, \postpnp{which can lead to} mismatches of up to $0.1$ in otherwise-identical waveforms~\citep{Islam:2021:meananomaly} assuming a white noise power spectral density.
It is not clear how the mismatch changes for realistic detector noise.

Our inferences of the eccentricities of our sources may be biased by neglecting this parameter, though, it is difficult to ascertain how this systematic error compares to other imperfections in the waveform model.
Investigations into the extent of this bias are ongoing.
However, the waveform amplitude modulations caused by orbital eccentricity appear to be qualitatively different than the changes induced by the mean anomaly. 
Hence, we suspect that our conclusions are relatively insensitive to this parameter.

An additional parameter that is fixed within \texttt{SEOBNRE} is the value of the spin-induced precession parameter, $\chi_{\rm p}${~\citep{Hannam:2014:chi_p, Schmidt:2015:chi_p}}.
While we can sample over the component aligned spins $\chi_1$ and $\chi_2$, we cannot probe misaligned spins with \texttt{SEOBNRE}{, enforcing an assumption that $\chi_{\rm p} = 0$}.
Precession has been shown to mimic the effects of eccentricity in gravitational waveforms for high-mass systems like GW190521A~\citep{Romero-Shaw:2020:GW190521, JuanHeadOn}.
Efficient waveform models than include the effects of both spin-induced precession and eccentricity are not yet available, so we are not currently able to measure both parameters simultaneously.

Reweighting is increasingly inefficient for low-mass events, i.e., those that require data segments with durations $D>\unit[4]{s}$.
With more cycles contained in longer-duration waveforms, systematic discrepancies between our proposal (quasi-circular) model \texttt{IMRPhenomD} and our target (eccentric) model \texttt{SEOBNRE} build up, manifesting in larger differences between the proposal and target likelihoods{; see Figure~\ref{fig:overlaps} in Appendix~\ref{sec:overlapApp} for a demonstration of the overlap between the two waveforms decreasing as source mass decreases, increasing the number of cycles in-band}.
There are two neutron star-black hole (NSBH) merger candidates in GWTC-2, with $D=\unit[16]{s}$ (GW190814A) and $D=\unit[64]{s}$ (GW190426A).
There are three other events with $D=\unit[16]{s}$ (GW190527A, GW190728A and GW190924A) and nine events with $D=8$~s (GW190412A, GW190512A, GW190630A, GWS190707A, GW190708A, GW190720A, GW190803A, GW190828A and GW190930A).\footnote{{We analyse segment durations that match those used in GWTC-2~\citep{GWTC-2}. An eccentric binary inspirals more rapidly than a non-eccentric binary with the same parameters. For a given orbital period, an eccentric binary is closer at periapsis than it would be in a circular orbit, increasing the energy that is therefore lost to gravitational radiation. Proposed eccentric waveforms are thus shorter than quasi-circular waveforms with otherwise identical parameters, so all waveforms that can be drawn from the eccentricity prior are therefore within the segment duration deemed adequate for quasi-circular parameter estimation.}}
Reweighting samples for most of these long-duration events is currently computationally impractical.

Low-mass black holes are less likely to merge via the gravitational-wave capture events that lead to eccentricities approaching unity in dense cluster environments~\citep[see, e.g.,][]{GondanKocsis2021}.
Higher-mass binaries with masses close to the pair-instability mass gap are also more likely to contain components that have formed hierarchically in a dynamical environment.
We therefore exclude ten low-mass BBH events and two NSBH candidates from this work.
We anticipate that it will be possible to analyse these events with new eccentric waveforms that are efficient enough to use for direct parameter estimation, so defer this analysis to future work.

\section{Results}\label{sec:results}

\begin{figure*}
    \centering
    \includegraphics[width=0.6\textwidth]{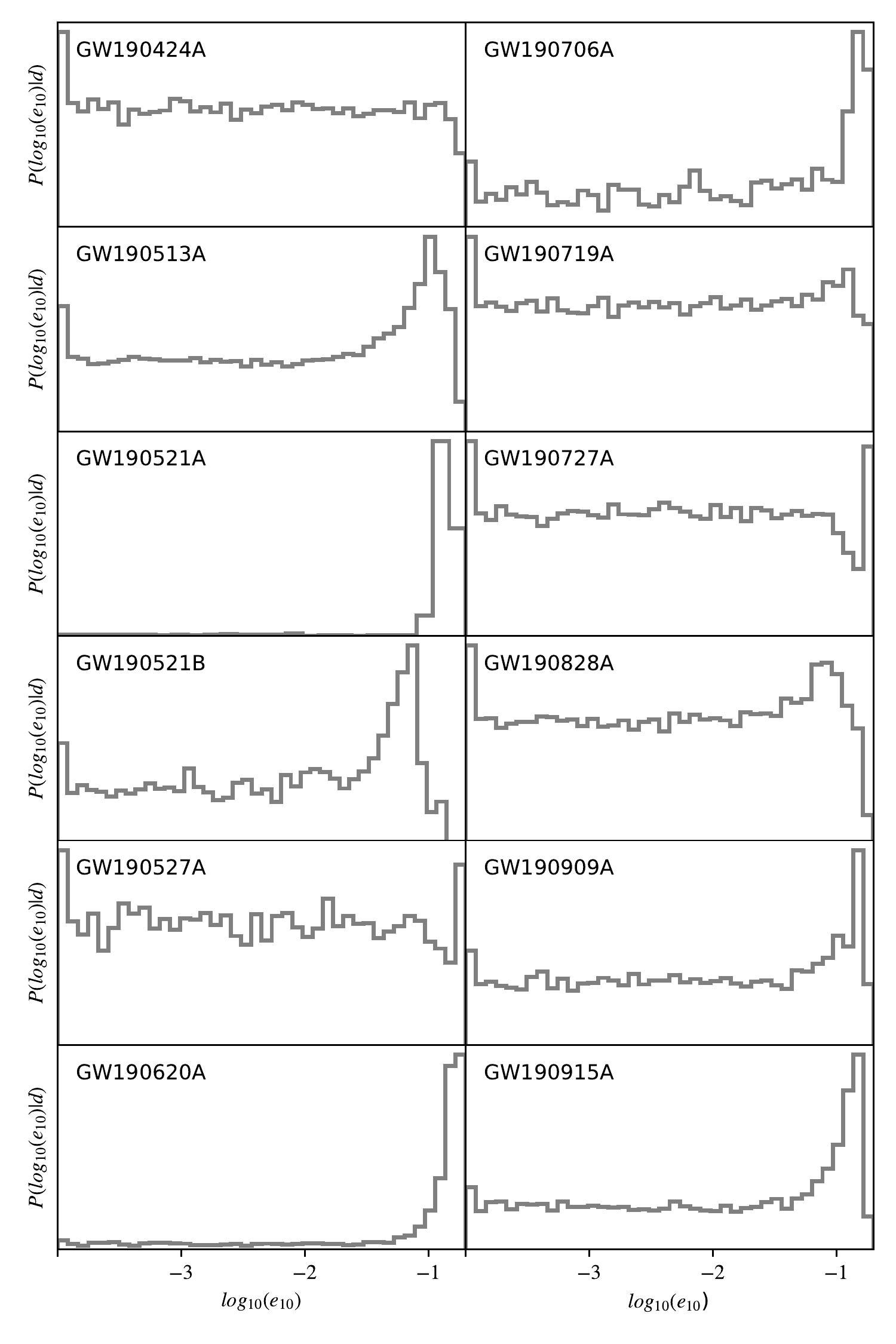}
    \caption{Posterior probability distributions on $e_{10}$ for the 12 events in GWTC-2 with eccentricity posteriors that have the most support for eccentricity $e_{10} \geq 0.05$.}
    \label{fig:interesting_ecc}
\end{figure*}

\begin{figure}
    \centering
    \includegraphics[width=0.45\textwidth]{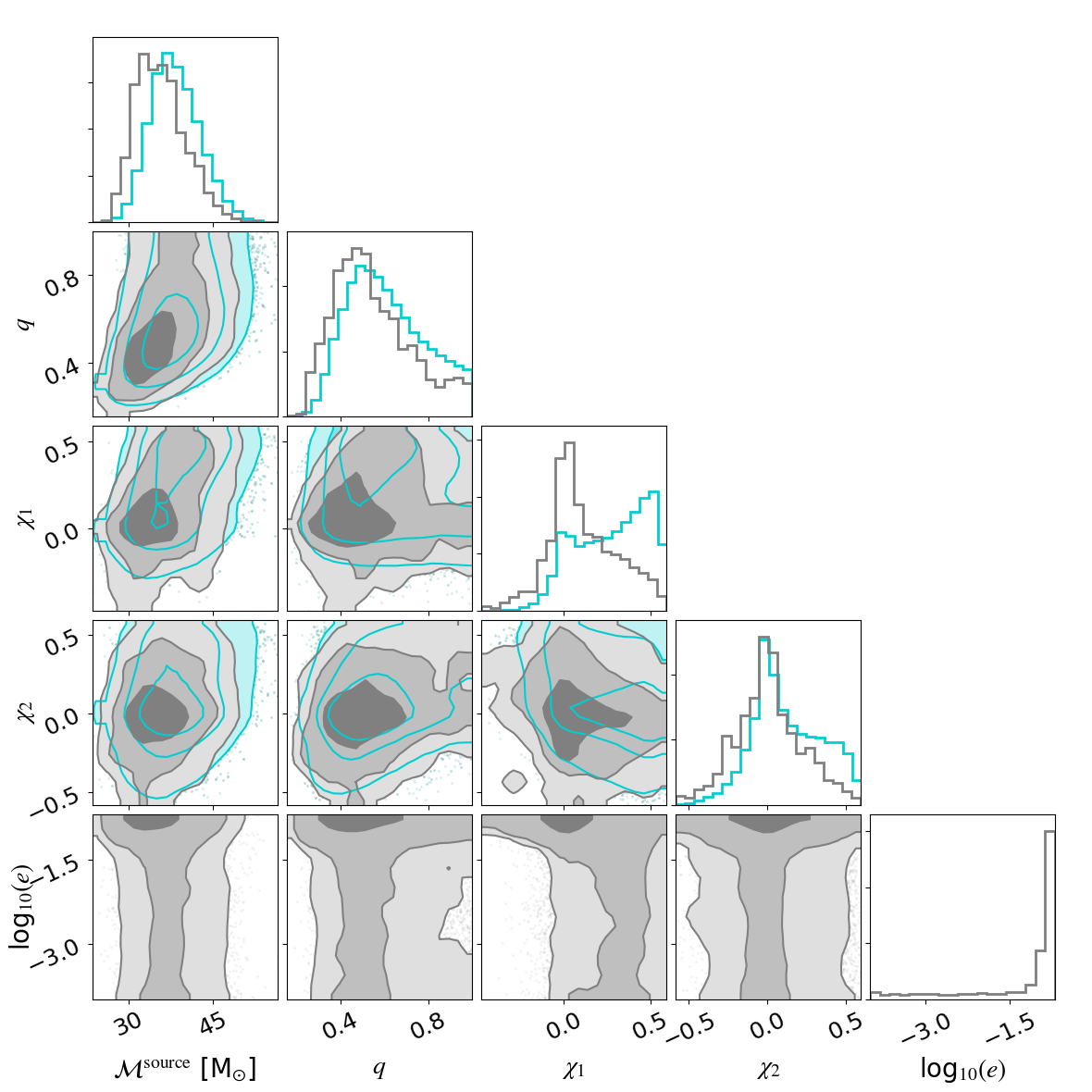}
    \caption{Posterior probability distributions on intrinsic parameters for GW190620A, with proposal (circular) parameter estimation results shown in teal and reweighted eccentric posteriors shown in grey. There is a slight visible correlation between source-frame chirp mass and eccentricity, as well as mass ratio and eccentricity. There is a clearer correlation between the aligned spin of the primary, $\chi_1$, and eccentricity.}
    \label{fig:GW190620A_intrinsic}
\end{figure}

\begin{figure}
    \centering
    \includegraphics[width=0.45\textwidth]{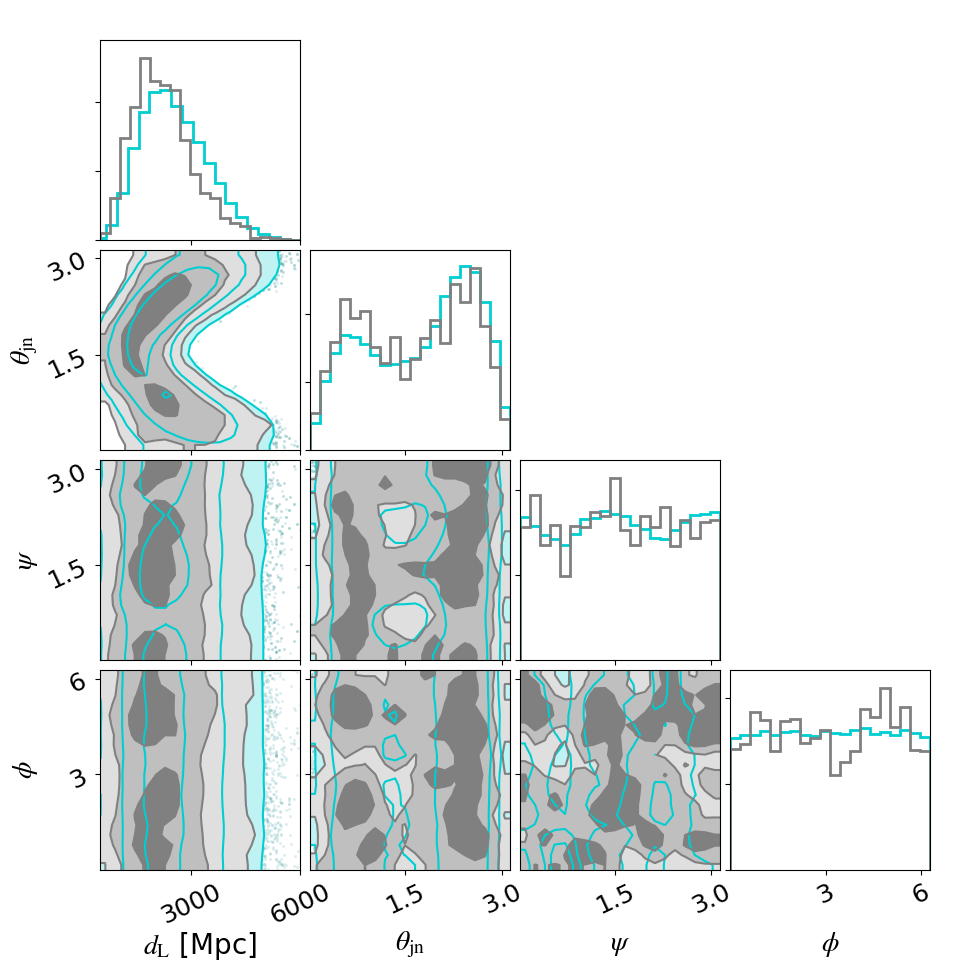}
    \caption{Posterior probability distributions on extrinsic parameters for GW190620A. The eccentric posterior causes a slight shift in the posterior to lower luminosity distances.}
    \label{fig:GW190620A_extrinsic}
\end{figure}

\begin{figure}
    \centering
    \includegraphics[width=0.5\textwidth]{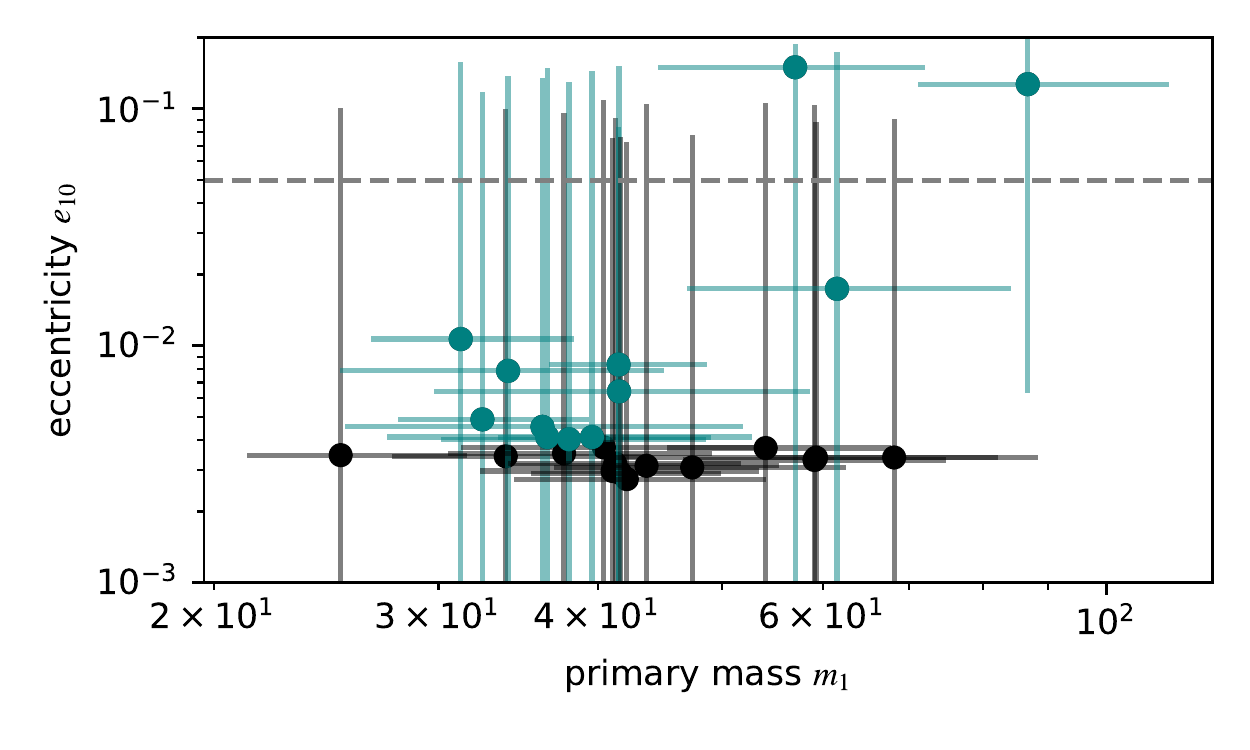}
    \caption{Scatter plot of source-frame primary mass $m_1$ against $e_{10}$ for the 26 BBH newly analysed in this paper, with error bars showing the $90\%$ credible range of the posterior across both axes and a dashed grey line at $e_{10} = 0.05$. Events with points above the grey dashed line are GW190620A and GW190521A, which are two of the highest-mass events in GWTC-2. Events highlighted in teal are those plotted in Figure \ref{fig:interesting_ecc} and tabulated in Table \ref{tab:Bayes_and_percentages}.}
    \label{fig:m1_vs_ecc}
\end{figure}

In Figure \ref{fig:eccentricityviolin}, we display the posterior probability distributions for eccentricity at $10$~Hz, $e_{10}$, for all of the binary black hole systems so far analysed for eccentricity with \texttt{SEOBNRE}. 
Corner plots containing fully- and partially-marginalised single- and double-dimensional posterior probability distributions for all other waveform parameters are available online for all events.\footnote{\href{https://github.com/IsobelMarguarethe/eccentric-GWTC-2/tree/main/seobnre}{github.com/IsobelMarguarethe/eccentric-GWTC-2/seobnre}}
{Consistent with \citet{Payne2019}, we consider sampling efficiency $> 1\%$ to be adequate.}
{The number of effective samples in the eccentric posterior after reweighting is $>500$ for all events presented here, with an average of $17,477$, a maximum of $54,395$ (GW190413B) and a minimum of $541$ (GW190521A).
The average reweighting efficiency is $45\%$, with a maximum of $90\%$ (GW190731A) and a minimum of $2\%$ (GW190521A and GW190803A).} 
new{The reweighting efficiency is particularly low for GW190521A because we also reweight from the old eccentricity prior to the new eccentricity prior; before doing this, the number of samples is 726.}
There are 12 events with marginalised eccentricity posteriors that show support for eccentricity $e_{10}\geq0.05$. 
We display these posteriors in Figure \ref{fig:interesting_ecc}.
The eccentricity posteriors for all other events are provided in {Appendix~\ref{sec:quasicircularApp}}.

\subsection{Events with $e_{10} \geq 0.05$}

There are two events that have more than $50\%$ of their posterior probability distribution above $e_{10} \geq 0.05$: GW190521A and GW190620A. 
There are also ten events that have support for $e_{10} \geq 0.05$ while remaining consistent with having negligible eccentricity. 
Of these ten events, three have eccentricity posteriors peaking in the range $0.1 \leq e_{10}$ and three have eccentricity posteriors peaking in the range $0.05 \leq e_{10} \leq 0.1$.
We provide the percentages of the eccentricity posterior above $0.1$ and $0.05$ for the 12 events of interest in Table \ref{tab:Bayes_and_percentages}, in addition to the natural-log Bayes factors $\ln \mathcal{B}$ for the hypotheses that $e_{10} \geq 0.1$ ($0.05$) against the hypothesis that $e_{10} < 0.1$ ($0.05$).
We display the posterior probability distribution for the eccentricity of these 12 events in Figure \ref{fig:interesting_ecc}.

\postpnp{In a sufficiently large population of entirely circular binaries, some events will appear to have non-zero eccentricity due to random fluctuations.}
In order to provide a different perspective on the statistical significance for eccentricity in GW190521A and GW190620A, we also calculate a frequentist $p$-value testing the hypothesis that the data are described by the \texttt{SEOBNRE} waveform with an eccentricity value of $e_{10}=0$.
We find that the frequentist confidence intervals for $e_{10}$ exclude $e_{10}=0$ with $\gtrsim99.9\%$ confidence (see Figure \ref{fig:all_eccentric_likelihoods} in Appendix~\ref{sec:likelihoodApp}).
This high statistical confidence illustrates that the eccentricity we observe is not due to random fluctuations amplified by trial factors.
Of course, this test does not tell us if the observed eccentricity is actually due to covariance with relativistic precession or other systematic error in the \texttt{SEOBNRE} waveform, a topic we return to below.

\subsubsection{GW190620A}

The eccentricity posterior for GW190620A has $e_{10} \geq 0.05$ at  $74\%$ confidence{, and contains 1269 samples after reweighting with an efficiency of $10\%$}.
The hypothesis that GW190620A has $e_{10} \geq 0.05$ is preferred to the hypothesis that $e_{10} < 0.05$ with $\ln\mathcal{B} = 2.48$.
GW190620A is a moderately high-mass binary with a total mass $\approx 92$~M$_\odot$ in the source-frame. 

While GW190521A was found by the LIGO-Virgo analysis to have strong support for in-plane spin~\citep{GW190521-detection}, the {LIGO-Virgo posterior distribution for the GW190620A value of $\chi_{\rm p}$ was uninformative, with little significant deviation from the prior}.
However, the posterior probability for effective aligned spin $\chi_{\rm eff}$ is found to peak at $\sim 0.3$, consistent with the \texttt{IMRPhenomD} posterior, as shown in Figure \ref{fig:GW190620A_intrinsic}. 
In contrast to the quasi-circular \texttt{IMRPhenomD} analysis, the eccentric \texttt{SEOBNRE} posterior {probability distributions for component spins $\chi_1$ and $\chi_2$ more closely resemble the prior, with both distributions unimodally peaked at $0$ and showing lower support for moderate positive spin}.
The eccentric posterior also has a slight preference for lower masses and a more extreme mass ratio, as shown in Figure \ref{fig:GW190620A_intrinsic}, and a slightly lower distance, as shown in Figure \ref{fig:GW190620A_extrinsic}.

\subsubsection{GW190521A}
We also provide updated statistics for GW190521A using the revised prior for $e_{10}$.
These results are qualitatively similar to previously published analyses.
The eccentricity posterior for GW190521A has $e_{10} \geq 0.1$ at greater than $92\%$ confidence, and $e_{10} \geq 0.05$ at greater than $93\%$ confidence.
The hypothesis that GW190521A has $e_{10} \geq 0.05$ is preferred to the hypothesis that $e_{10} < 0.05$ with a natural-log Bayes factor $\ln\mathcal{B} = 3.90$.
{Since the eccentric posterior for GW190521A contains the fewest samples of all events, we confirm our eccentricity measurement by performing massively parallel inference with \texttt{parallel\_bilby}~\citep{ParallelBilby}, splitting our analysis with \texttt{SEOBNRE} over 800 CPUs.
We restrict the chirp mass, component mass and spin priors to reduce the time required for such a computationally demanding endeavour.
The posterior probability distribution obtained with direct sampling is consistent with that obtained with reweighting, and can be found in Appendix \ref{app:directsampling}, along with further details about that analysis.}

\subsection{A correlation between primary mass and eccentricity?}

We speculate that eccentricity might be observed preferentially in high-mass systems.
In Figure \ref{fig:m1_vs_ecc}, we plot the median source-frame primary mass and median eccentricity of each event, with bars extending over the $90\%$ credible range of each parameter.
{Source-frame masses are obtained assuming a flat $\Lambda$CDM universe with cosmological parameters $H_0=\unit[67.7]{km s^{-1} Mpc^{-1}}$ and $\Omega_0=0.307$ as defined in \citet{Planck2015}.}
The two BBH events with signatures of eccentricity are both associated with large primary mass.
If this correlation is real, it might provide clues as to the origin of eccentric mergers.
Of course, the correlation could also be indicative of systematic error: gravitational waveform analysis is more sensitive to merger physics when the signal is short, as it is for high-mass BBH, and imperfections in the waveform are likely to be most pronounced in this regime.

\subsection{Correlation between spin / precession and eccentricity}

GW190521A has previously been shown to be consistent with both an eccentric and a spin-precessing system~\citep{GW190521-detection, GW190521-implications, Romero-Shaw:2020:GW190521,Gayathri2020}.
GW190620A does not have strong evidence for precession~\citep{GWTC-2}, but is found by our quasi-circular analysis to support a non-zero value of the effective inspiral spin parameter, $\chi_{\rm eff} \sim 0.3$ \citep{Kidder}.
However, when we reweight to our target (eccentric) posterior, higher values of $\chi_1$ and $\chi_2$ are weighted lowly, giving us $\chi_{\rm eff} = 0.06^{+ 0.2}_{- 0.2}$ after reweighting.
There is a clear correlation between $\chi_1$ and eccentricity in the central-lower panel of Figure \ref{fig:GW190620A_intrinsic}; this agrees with the correlation between effective spin and eccentricity noted by \cite{OSheaKumar2021}.
Our findings for GW190620A support the argument that eccentric systems may be mistaken by quasi-circular parameter estimation efforts as systems with non-zero aligned spin.

\section{Discussion}
\label{sec:discussion}

Since the fraction of binary black holes merging with detectable eccentricity in dense star clusters is thought to be robust to changes in simulation parameters, observations of orbital eccentricity can be used to constrain the fraction of LIGO--Virgo binaries being produced in these environments.
In \citet{Zevin:2021:seleccentricity}, the lower limit on this branching fraction, $\beta_{\rm c}$, is shown to be $0.14$ ($0.27$) at $95\%$ \postpnp{credibility} for a number of observations with $e_{10} \geq 0.05$, $N_{\rm ecc} = 1$ ($2$), when the total number of observed BBHs is $N_{\rm obs} = 46$.

In this work, we present GW190620A, a source with $74\%$ of its eccentricity posterior above $e_{10} = 0.05$.
Combining this event with GW190521, there are now two gravitational-wave events with signatures of non-zero eccentricity.
We include measurements for 36 BBH in this work, but use N$_{\rm obs} = 46$ to calculate conservative lower limits on the cluster branching fraction.
With N$_{\rm ecc} = 2$, the cluster branching fraction $\beta_{\rm c} \geq 0.27$. 
If GW190521A is actually a quasi-circular precessing system and GW190620A is truly eccentric, then $\beta_{\rm c} \geq 0.14$

While we highlight the two events with the majority of their posterior support at $e_{10} \geq 0.05$, there are an additional ten events that show support for eccentricity, remaining consistent with or peaking at $e_{10} \gtrsim \mathcal{O} (0.01)$.
Although these events have less statistically significant support for eccentricity, with no more than $38\%$ of their posterior probability in the region of $e_{10} \geq 0.05$, their support relative to other GWTC-2 events (see Table \ref{tab:Bayes_and_percentages_not_ecc}) introduces the possibility that we may have $\geq 4$ eccentric events in GWTC-2.
If these events truly are eccentric---not just statistical fluctuations, or capturing the effects of spin-induced precession---then dense star clusters alone cannot account for the abundance of eccentric binaries~\citep{Zevin:2021:seleccentricity}.
This would mean that other channels capable of producing eccentric compact binaries must be contributing significant quantities of mergers to our catalogues.
Recent work has shown that in environments like active galactic nuclei discs, up to $\sim 70\%$ of binary black hole mergers retain detectable eccentricity within the LIGO--Virgo band~\citep{Samsing:2020:AGN, Tagawa:2021:AGN}, depending on the freedoms of motion available to binaries within the disc.
While we do not yet well-understand active galactic nuclei as dynamical formation environments, a spurious overabundance of eccentric mergers may, in fact, indicate that alternative dynamical environments, such as active galactic nuclei discs, play a significant role in producing mergers detected by LIGO and Virgo.

Eccentric waveform model development is ongoing, and recent models are becoming efficient enough to perform parameter estimation directly~\citep[e.g.,][]{TEOBResumS, Islam:2021:meananomaly, SEOBNREfrequencydomain, Setyawati2021}.
{Additionally, model-independent analyses such as that simulated in ~\citet{Dalya:2021:BW_reconstructed_signals} may be useful for future discovery of high-eccentricity sources, which can be missed by searches that assume quasi-circular signals~\citep[e.g.,][]{Brown09}}.
It is not computationally feasible to analyse tens of long-duration events with \texttt{SEOBNRE}, but we anticipate that it will soon be possible to compute eccentric analysis of catalogues using new, inexpensive waveform models.
Different waveform model families are based on different physical approximations, and different eccentric waveform models may use different definitions of eccentricity; any future studies comparing analyses with multiple models must quantify the effects of these differences. 
Additionally, while there are no waveform models currently available that contain a variable mean anomaly, the effects of eccentricity and the effects of spin-induced precession, we hope that waveform development in this direction~\citep[e.g.,][]{Klein2021} will enable us to disentangle of the effects of these three parameters in future work.

\section{Acknowledgements}
We thank Mike Zevin for useful discussions and comments on the manuscript{, and thank our anonymous referee for careful reading and insightful suggestions}.
This work is supported through Australian Research Council (ARC) Future Fellowships FT150100281, FT160100112, Centre of Excellence CE170100004, and Discovery Project DP180103155.
Computing was performed using the LIGO Laboratory computing cluster at California Institute of Technology, supported by National Science Foundation Grants PHY-0757058 and PHY-0823459{, and the OzSTAR Australian national facility at Swinburne University of Technology, which receives funding in part from the Astronomy National Collaborative Research Infrastructure Strategy (NCRIS) allocation provided by the Australian Government.}
LIGO was constructed by the California Institute of Technology and Massachusetts Institute of Technology with funding from the National Science Foundation and operates under cooperative agreement PHY-1764464. Virgo is funded by the French Centre National de Recherche Scientifique (CNRS), the Italian Istituto Nazionale della Fisica Nucleare (INFN) and the Dutch Nikhef, with contributions by Polish and Hungarian institutes.

\bibliography{bib}

\appendix
\section{Events consistent with quasi-circularity}\label{sec:quasicircularApp}

\begin{table*}
\centering
\caption{Percentages of the eccentricity posterior probability distribution above $0.1$ and $0.05$ for the 14 events analysed in this paper that have low support for $e_{10} \geq 0.05$. We also provide the natural log Bayes factors $\ln \mathcal{B}$ for the hypotheses that $e_{10} \geq 0.1$ ($0.05$) against the hypothesis that $e_{10} \leq 0.1$ ($0.05$). These events all have less than $16\%$ of their posterior above $e_{10} = 0.05$, and have $\ln \mathcal{B}(e_{10} \geq 0.05) \leq -0.2$.
\label{tab:Bayes_and_percentages_not_ecc}}
\bgroup
\def\arraystretch{1.5}
\begin{tabular}{c|c|c|c|c|c}
Event name & percentage $e_{10} \geq 0.1$ & percentage $e_{10} \geq 0.05$ & $\ln \mathcal{B} (e_{10} \geq 0.1)$  & $\ln \mathcal{B} (e_{10} \geq 0.05)$ & reweighting efficiency (\%) \\
\hline
GW190408A & $4.86$ & $13.79$ & $-0.69$ & $-0.35$  & 48 \\
GW190413A & $2.17$ & $9.84$ & $-1.24$ & $-0.65$  & 70 \\
GW190413B & $4.73$ & $13.49$ & $-0.68$ & $-0.35$ &  88 \\ 
GW190421A & $1.58$ & $9.58$ & $-1.81$ & $-0.75$  & 79 \\ 
GW190503A & $3.67$ & $11.78$ & $-0.98$ & $-0.51$ &  61 \\ 
GW190514A & $5.83$ & $14.81$ & $-0.45$ & $-0.24$  & 85 \\
GW190517A & $5.38$ & $13.04$ & $-0.52$ & $-0.34$  & 4 \\
GW190519A & $5.08$ & $14.95$ & $-0.59$ & $-0.20$  & 27 \\ 
GW190602A & $3.85$ & $12.27$ & $-0.84$ & $-0.43$  & 54 \\ 
GW190701A & $5.64$ & $15.30$ & $-0.50$ & $-0.20$  & 84 \\
GW190731A & $2.21$ & $9.71$ & $-1.10$ & $-0.55$  & 90 \\ 
GW190803A & $4.08$ & $11.65$ & $-0.99$ & $-0.58$  & 2 \\ 
GW190910A & $1.32$ & $10.04$ & $-1.20$ & $-0.47$ & 63 \\  
GW190929A & $3.28$ & $12.91$ & $-0.76$ & $-0.30$  & 48 \\ 
\end{tabular}
\egroup
\end{table*}

\begin{figure*}
    \centering
    \includegraphics[width=0.6\textwidth]{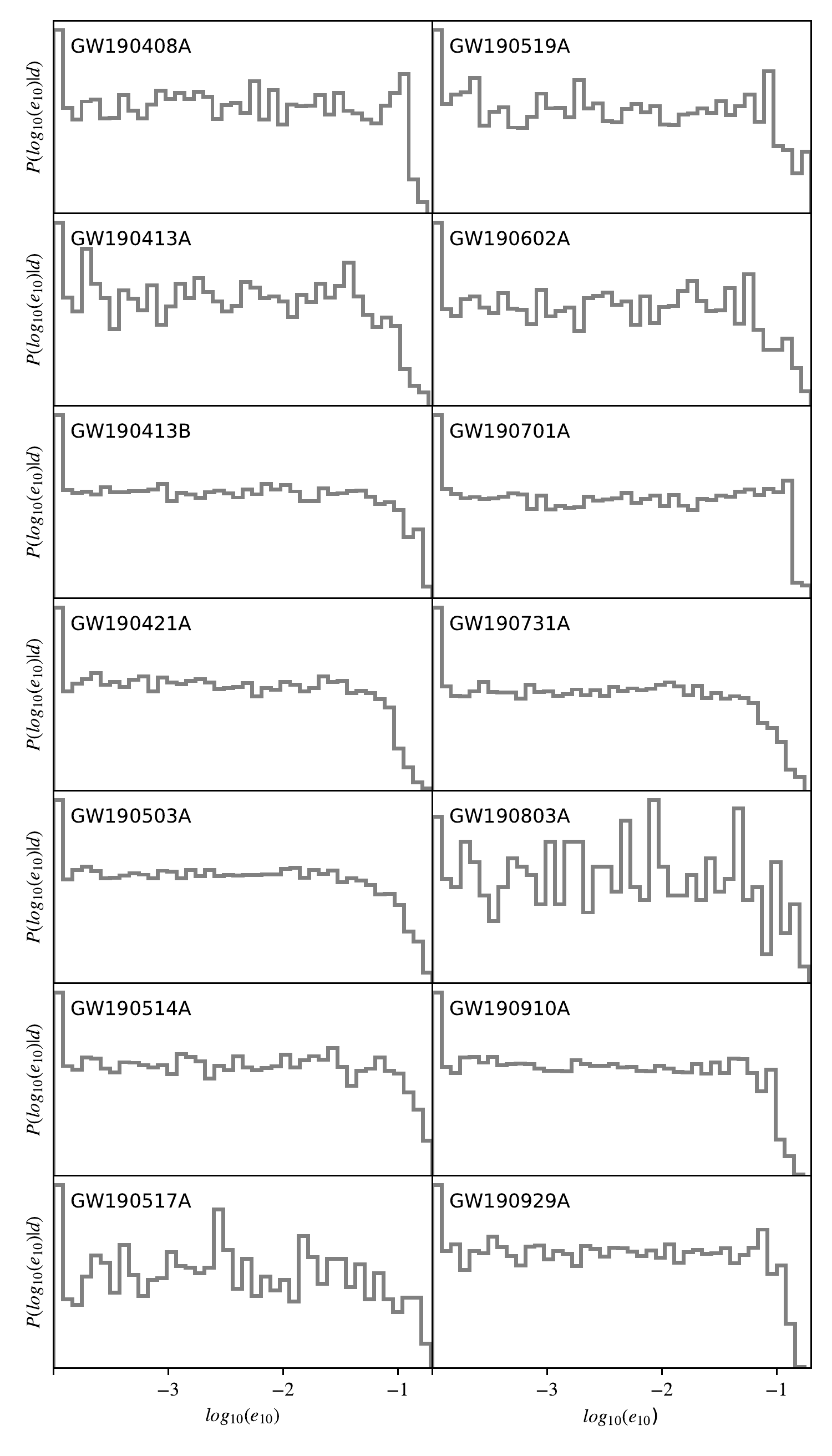}
    \caption{Posterior probability distributions on $e_{10}$ for 14 events in GWTC-2 with eccentricity posteriors that have little support for $e_{10} \geq 0.05$.}
    \label{fig:not_interesting_ecc}
\end{figure*}

We provide the percentages of the posterior above $e_{10}=0.05$ and $0.1$ in Table \ref{tab:Bayes_and_percentages_not_ecc} for events that do not have significant posterior support for $e_{10} \geq 0.05$.
All of these events have less than $16\%$ of their posterior support above $e_{10} = 0.05$, so are consistent with quasi-circularity within our sensitivity limits to eccentricity.
We also provide here the natural-log Bayes factors for the hypotheses that $e_{10} \geq 0.05$ and $0.1$.
All of these events have $ln\mathcal{B} \leq -0.2$ for the hypothesis that $e_{10} \geq 0.05$ relative to the hypothesis that $e_{10} \leq 0.05$, implying that the data does not favour the eccentric hypothesis over the quasi-circular hypothesis. 
We show the posterior probability distributions for the eccentricity of these events in Figure \ref{fig:not_interesting_ecc}.

\section{Eccentric likelihood / eccentric posterior with uniform prior}\label{sec:likelihoodApp}

\begin{figure*}
    \centering
    \includegraphics[width=0.75\textwidth]{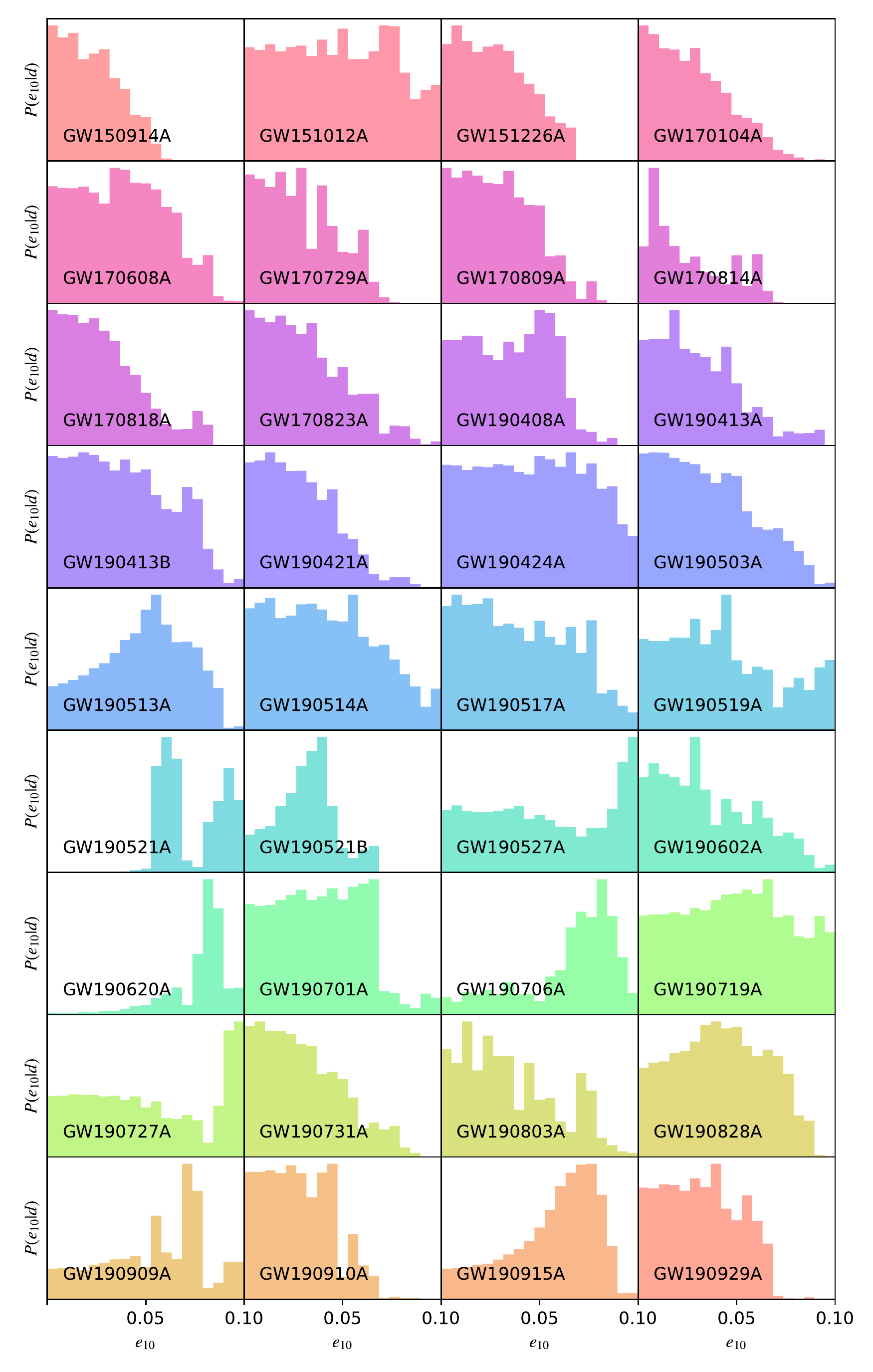}
    \caption{The {posterior probability distributions under a uniform eccentricity prior} for all 36 BBH events so far analysed for eccentricity using \texttt{SEOBNRE}. {This is equivalent to the likelihood distribution used in our primary analysis using a log-uniform prior on eccentricity.}}
    \label{fig:all_eccentric_likelihoods}
\end{figure*}

We plot the eccentric model likelihood for all 36 BBH \postpnp{so far analysed for eccentricity} in GWTC-2 in Figure \ref{fig:all_eccentric_likelihoods}.
The eccentric likelihood is obtained by dividing out the log-uniform prior on eccentricity from the eccentric posterior distribution. 
The resulting likelihood is equivalent to the posterior that would be obtained if we used a uniform sampling prior on $e_{10}$.
While the log-uniform prior represents our prior expectations of the eccentricity of our sources, dividing this out better illustrates which events are not well-supported by the {negligible eccentricity} hypothesis.
GW190521A and GW190620A are the only two events with negligible likelihood amplitude at  $e_{10}=10^{-4}$.

\newpage
\new{
\section{Overlap between \texttt{SEOBNRE} and \texttt{IMRPhenomD}, and the mass dependence of the upper eccentricity constraint}\label{sec:overlapApp}
}

\begin{figure}
    \centering
    \includegraphics[width=0.5\textwidth]{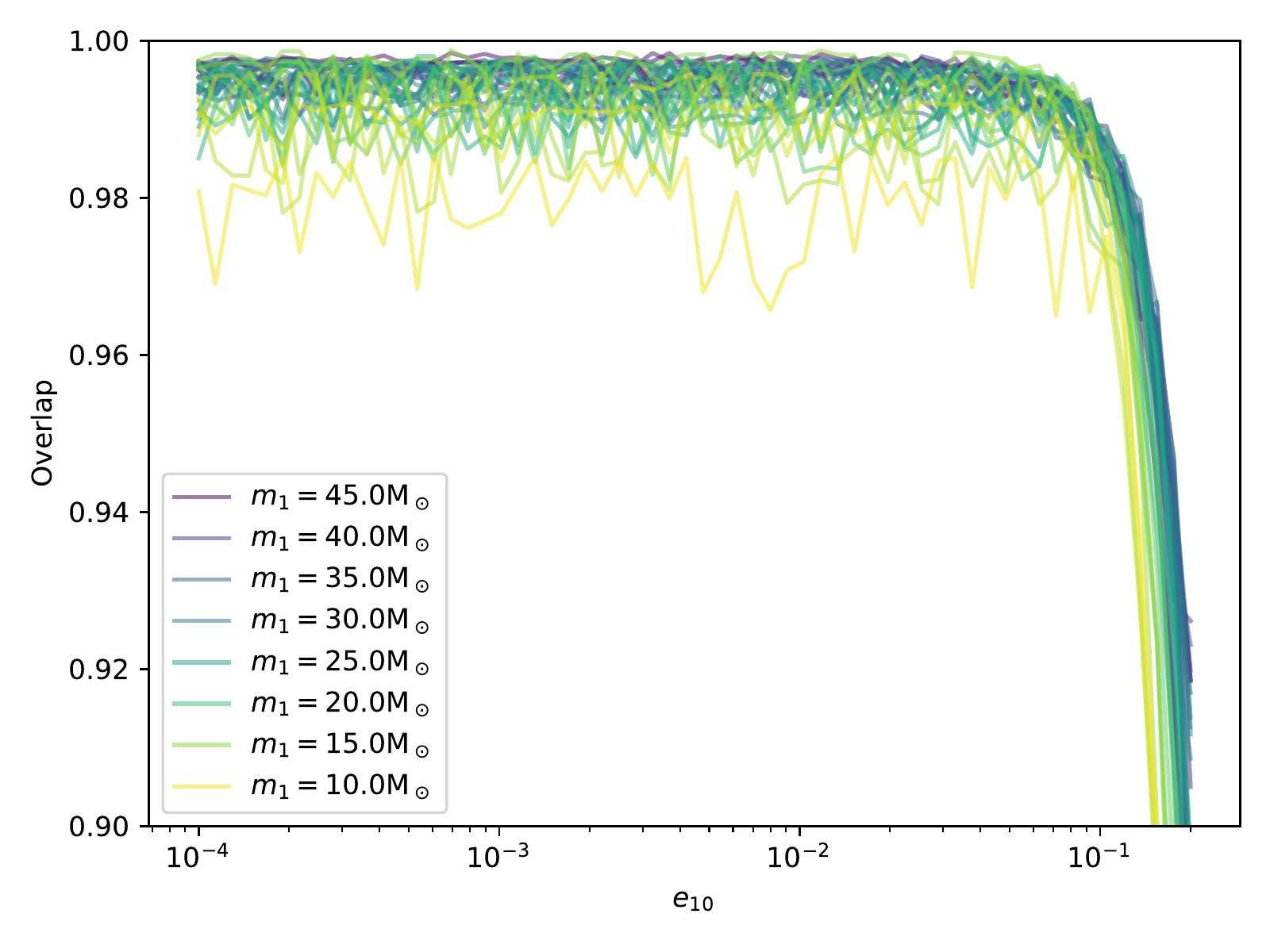}
    \caption{\label{fig:overlaps}\new{The overlap between \texttt{SEOBNRE} and \texttt{IMRPhenomD} with identical parameters but with eccentricity in the \texttt{SEOBNRE} waveform. We plot the overlap curves for systems with $q=0.8$ and detector-frame $m_1$ from \unit[$10$]{M$_\odot$} to \unit[$45$]{M$_\odot$} at intervals of \unit[$1$]{M$_\odot$}, with legend labels at every \unit[$5$]{M$_\odot$} interval. We use a duration of \unit[4]{s} and sampling frequency of \unit[4096]{Hz}. Because the mismatch between two waveforms tends to worsen as the number of cycles in-band increases, the maximum overlap gets lower as the mass of the system decreases, leading to lower reweighting efficiency for lower-mass systems. However, lower-mass systems also deviate from semi-constant overlap at lower eccentricities, so we are able to constrain their eccentricity to lower values.}}
    \label{fig:mass-dependence-overlap}
\end{figure}

\new{
We observe that higher-mass systems have higher credible limits on their minimum eccentricity at \unit[10]{Hz} than lower-mass systems. 
It is easier to constrain the eccentricity of lower-mass systems because they have more cycles in-band than higher-mass systems, so more of the eccentricity-imprinted inspiral is observed.
In Figure \ref{fig:mass-dependence-overlap}, we plot the overlap between \texttt{SEOBNRE} and \texttt{IMRPhenomD} as the eccentricity encoded in the \texttt{SEOBNRE} waveform is increased.\footnote{See \citet{Lower18} for details of the overlap calculation. For this demonstration we use just one detector with LIGO Livingston-like sensitivity.}
Where the overlap is roughly constant (with oscillations due to hard-coded changes in the mean anomaly of the eccentric waveform, which we cannot change), the eccentric and quasi-circular waveform are indistinguishable at current detector sensitivity. 
Above some value of eccentricity, the overlap between \texttt{SEOBNRE} and \texttt{IMRPhenomD} rapidly decreases.
The value of eccentricity at which this happens is the lower limit of eccentricity sensitivity for that particular waveform.
This means that, for lower-mass systems, it should be possible to measure smaller eccentricities than for higher-mass systems.
}

\begin{figure*}
    \centering
    \includegraphics[width=0.45\textwidth]{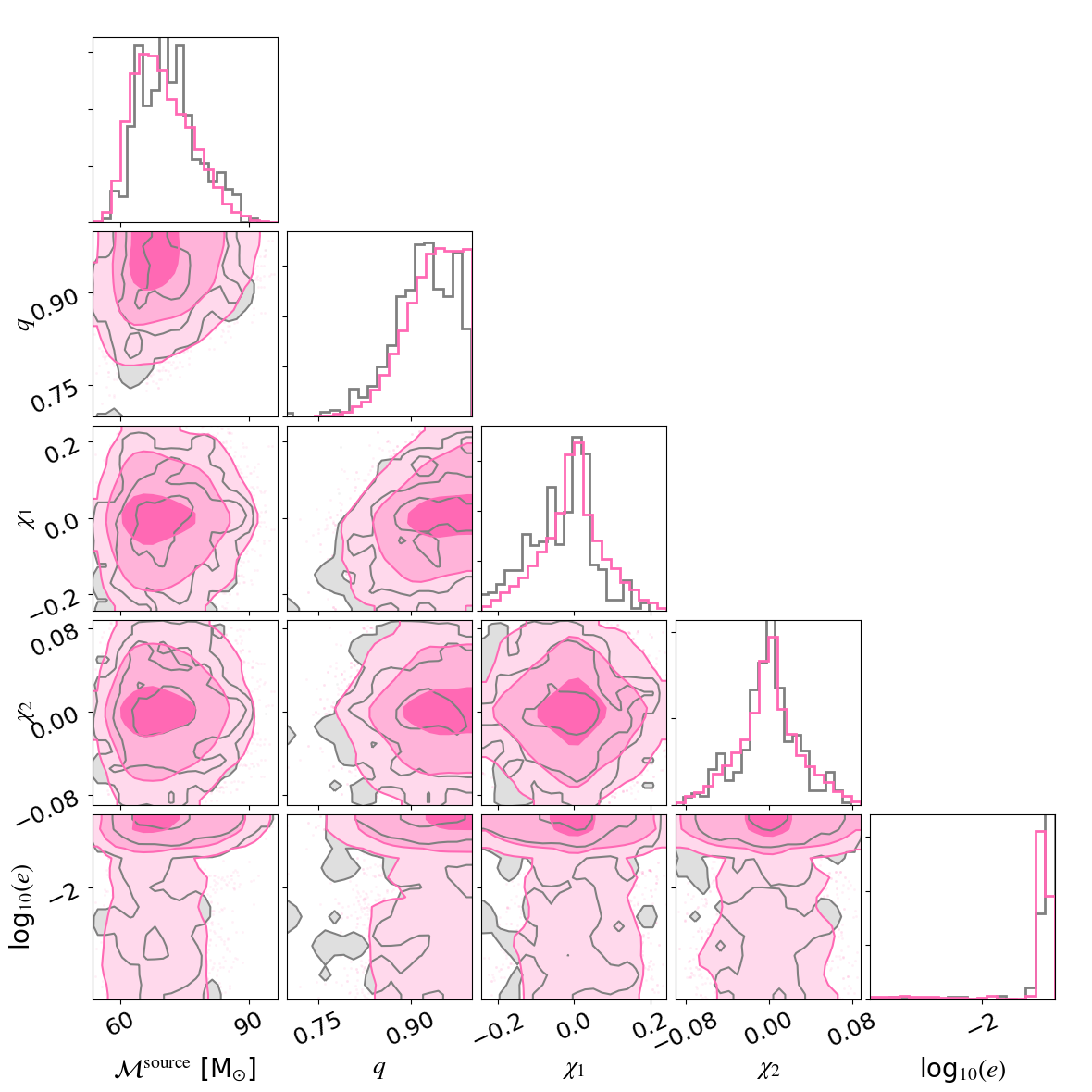}
    \caption{{Posterior probability distributions on intrinsic parameters for GW190521A, with reweighted results shown in grey and directly sampled results shown in pink.\label{fig:GW190521A}}}
\end{figure*}

\newpage
{
\section{Massively parallel analysis to confirm eccentric posteriors with direct sampling}\label{app:directsampling}
}
{
To confirm that our reweighted eccentricity posteriors are consistent with those obtained with direct sampling, we use \texttt{parallel\_bilby}~\citep{ParallelBilby} to directly sample the posterior of GW190521A with eccentric waveform model \texttt{SEOBNRE} using 800 parallel cores. 
Even with a large number of cores, the full analysis is computationally prohibitive, so we restrict our priors to a region in the vicinity of the posterior maximum: detector-frame chirp masses between $90$ and $140$~M$_\odot$, individual component masses between $40$ and $140$~M$_\odot$, and $|\chi_1|<0.5$ and $|\chi_2|<0.3$.\footnote{The restricted prior run required $\sim\unit[35]{hr}$ of wall time with 800 cores.}}
{The posterior obtained with direct sampling (pink) is compared to that obtained with reweighting under the same prior restrictions (grey) in Fig.~\ref{fig:GW190521A}.
The two posteriors display the same strong posterior support for eccentricity above $e_{10} = 0.1$ while producing qualitatively similar posterior distributions for the other parameters.
This check gives us confidence that the reweighting method is reliable.
While direct sampling is possible for GW190521A---a single, short-duration event, with restricted priors---this is not practical for other events.}

\end{document}